# Monte Carlo study of the random-field Ising model

M. E. J. Newman

*Cornell Theory Center, Cornell University,*
*Ithaca, NY 14853-3801, U.S.A.*

G. T. Barkema

*Institute for Advanced Study, Olden Lane,*
*Princeton, NJ 08540, U.S.A.*

Using a cluster-flipping Monte Carlo algorithm combined with a generalization of the histogram reweighting scheme of Ferrenberg and Swendsen, we have studied the equilibrium properties of the thermal random-field Ising model on a cubic lattice in three dimensions. We have equilibrated systems of $L \times L \times L$ spins, with values of $L$ up to 32, and for these systems the cluster-flipping method appears to a large extent to overcome the slow equilibration seen in single-spin-flip methods. From the results of our simulations we have extracted values for the critical exponents and the critical temperature and randomness of the model by finite size scaling. For the exponents we find $\nu = 1.02 \pm 0.06$, $\beta = 0.06 \pm 0.07$, $\gamma = 1.9 \pm 0.2$, and $\bar{\gamma} = 2.9 \pm 0.2$.

## I. INTRODUCTION

The random-field Ising model (RFIM) is one of the best-studied glassy magnetic models [1]. It has been the subject of considerable controversy over the last ten or fifteen years, particularly concerning the nature of its phase transitions. The model consists of Ising spins $s_i$ on a lattice, governed by the Hamiltonian

$$H = -J \sum_{\langle ij \rangle} s_i s_j - \sum_i h_i s_i. \qquad (1)$$

$J$ is an interaction constant which we take to be positive so that the model is ferromagnetic, and the variables $h_i$ are random fields, one on each lattice site, chosen independently from some probability distribution $P(h)$. A number of different choices for $P(h)$ have been considered in the literature. The most common is the Gaussian distribution

$$P(h) = \frac{1}{\sqrt{2\pi}\sigma} \exp\left[-\frac{h^2}{2\sigma^2}\right]. \qquad (2)$$

The width $\sigma$ of the distribution is referred to as the "randomness" of the model. The studies presented in this paper all use this Gaussian distribution of fields, although our methods are by no means restricted to Gaussian fields, and could just as easily be applied to any other distribution.

One of the leading areas of contention over the RFIM has been the question of whether the model undergoes a phase transition from a high-temperature paramagnetic phase to a low-temperature ferromagnetic one for any value of the randomness parameter $\sigma$. The supersymmetry arguments given by Parisi and co-workers [2] led to the concept of "dimensional reduction", which appeared to indicate that the critical behavior of the RFIM in $d$ dimensions at sufficiently low (but non-zero) randomness should be identical to that of the normal Ising model (equivalent to $\sigma = 0$) in dimension $d - 2$. This in turn indicated that the model should not have a phase transition at finite temperature in three dimensions or fewer. However, a completely different and rather simpler argument based on the "droplet" theory of domain wall energies in the ferromagnetic state [3] seemed to indicate that a transition should exist in three dimensions for finite temperature and randomness. This particular puzzle has now been largely solved following the work of Imbrie [4] and also of Bricmont and Kupiainen [5], who have given arguments demonstrating that there should indeed be a phase transition to a ferromagnetic state in three dimensions, provided the randomness is sufficiently small. (Whilst the direct application of dimensional reduction turns out to be incorrect, the related result that the coefficients of the $\epsilon$-expansions should be the same, term by term, for the RFIM in $d$ dimensions and the normal Ising model in $d - 2$ dimensions, may still turn out to be useful, in the study of the RFIM out of equilibrium at T=0 [6]—see Section V.)

One might imagine that general questions such as these concerning the very existence of phase transitions in the model might be answerable by experiment. No experiments have been performed on true experimental realizations of the RFIM itself, but it has been demonstrated that dilute antiferromagnets in uniform external field fall into the same universality class [7,8], and a number of experiments have been performed to investigate the phase transitions of these systems in three dimensions [9,10]. These experiments have proven very difficult however, and their results inconclusive, because of the extremely slow, glassy dynamics of the system. In coming to equilibrium, the domain walls in an RFIM system can "pin" on the random fields, producing energy barriers to equilibration which are temperature-independent. As the temperature of the system is lowered, the timescale for thermal activation over a barrier of height $B$ increases exponentially as $\exp(B/kT)$, and so one does not have to go to particularly low temperatures to find a regime in which the timescale for thermal depinning of a pinned site is far longer than the length of any experiment. Furthermore, the timescale for depinning the entire domain wall increases exponentially as some power of the length of the wall, since it requires the simultaneous depinning of pinned sites all along the wall [11]. Thus the size of the RFIM that we can successfully equilibrate on laboratory timescales is also strictly limited. (This is reflected in renormalization group and scaling analyses of the problem [11–14] in which the temperature appears as an irrelevant parameter at the $T = 0$ fixed point governing the phase transition. In other words, the larger the region we wish to turn over, the more inadequate is the thermal energy available to do it. These issues are discussed in more detail in Section II.)

An alternative approach to answering questions about the model is to perform a Monte Carlo simulation. A number of simulations have been attempted [15–18] using the traditional techniques which have proved very successful in the study of the normal Ising model [19]. Such techniques however are subject to exactly the same problems as experiments: the timescales for equilibration of the system grow exponentially with decreasing temperature and increasing system size, limiting studies to very small systems and rather short times. (Simulations have typically only run for two or three times the estimated correlation time of the system.) Also, because of the random nature of the model, it is neccesary to repeat any Monte Carlo calculation for many different realizations of the random fields and average over the results in order to get a good estimate of the mean properties of the model. Typically, one has to average over a few hundred such realizations, and this decreases the amount of computer time available for the simulation of each individual system. In this paper, we describe Monte Carlo calculations performed on the RFIM using a "cluster-flipping" dynamics, which is far less susceptible to pinning problems than the Metropolis single-spin-flip dynamics employed in previous studies. Our algorithm has allowed us to perform considerably longer simulations (measured in terms of the correlation time) than have previously been possible, with a corresponding increase in the accuracy of the results. In addition, we have implemented our algorithms on a parallel computer, which allows us to study large numbers of different realizations of the randomness simultaneously.

Although the questions discussed above concerning the existence of a phase transition in the model have largely now been settled, there are a number of fundamental issues still to be decided, many of which are well-suited to study by the Monte Carlo method. In particular, in this paper we investigate the critical properties of the model in three dimensions including the form of the phase diagram and the values of critical exponents defined at the transition. The paper is laid out as follows. In Section II we describe our cluster-flipping algorithm and contrast it with the algorithms used in previous studies. In Section III we derive the generalization of the histogram method which we have used to improve the statistical quality of our simulation data. In Section IV we describe the finite-size scaling analysis we have performed to extract figures for the critical properties of the model. Section V contains our results and discussion. In Section VI we give our conclusions.

## II. THE ALGORITHM

Previous Monte Carlo studies of the RFIM [17] have made use of the Metropolis algorithm [20], in which the attempted moves are the flips of single Ising spins. Unfortunately, such single-spin dynamics places very stringent limits on the size of the system that can be simulated, because the equilibration time $\tau$ becomes exponentially long as the temperature drops below $T_c$. To understand why this is so, consider first a normal Ising system which is being cooled below its ferromagnetic transition. The system condenses into domains of aligned spins separated by domain walls with surface tension which goes like the inverse $\xi^{-1}$ of the correlation length, and therefore grows with decreasing temperature below $T_c$. For an island of one spin-state—spin up, say—in a sea of predominantly down-pointing spins, the combination of this surface tension and the net convexity of the island causes the island to shrink steadily and eventually to vanish. This is the primary mechanism by which the system rids itself of domain walls, and thereby reduces its free energy. Now consider the same situation in the case of the random-field model. Again we have domains separated by walls with a certain surface tension. However, it is now possible for the domain walls to "pin" on large local fields, by which we mean that there can be one spin on the edge of a domain which has a large local field $h_i$ with which it is aligned. If we have single-spin-flip dynamics, like the Metropolis algorithm, then in order to move the domain wall, we have to flip this spin, which demands a lot of energy, and is therefore rather unlikely. Such pinned spins are able to combat the surface tension in the wall of an island and prevent the island from evaporating, thus slowing the equilibration of the system in the ferromagnetic phase. Furthermore, since the energy barrier $B$ for flipping one of these crucial pinned spins is a temperature-independent constant, the time taken to anneal away an island becomes exponentially long as $\exp(B/kT)$ as the temperature gets lower.

One solution to this problem is to look for a Monte Carlo algorithm which, instead of flipping single spins at each move, flips groups of spins, or "clusters". Such algorithms have previously been used in the simulation of, for example, the normal Ising model [21,22] to greatly speed up the diffusion of domain walls. In the random-field case, we would expect the improvement in speed to be even more dramatic.

The idea is that we construct an algorithm which finds islands of up-pointing spins stranded in seas of down-pointing ones (or *vice versa*) and flips them over as a whole, rather than flipping them spin by spin. Then, even if some of the spins on the borders of the island were pinned by their large local fields, if the energy of the system is lowered by flipping the entire island, the move will still have a high likelihood of occurrence.

A simple cluster-flipping algorithm for the RFIM has been proposed by Dotsenko, Selke, and Talapov [23]. This algorithm is a straightforward extension of the Wolff algorithm [21] for the normal Ising model in which a thermal cluster is grown by starting with one randomly chosen spin, and then adding similarly-oriented neighboring ones with a probability $1 - \exp(-2\beta J)$. In the case of the normal Ising model, all the spins in the cluster are then simply flipped over. It can be shown (see, for example, Ref. [24]) that this satisfies the two crucial requirements for Monte Carlo algorithms of (i) ergodicity (i.e., being capable of reaching any state of the system in a finite number of steps) and (ii) detailed balance, which in this case means that the ratio $T_{\lambda\mu}/T_{\mu\lambda}$ of the rates for transitions from state $\lambda$ to state $\mu$ and back again should equal the ratio $\exp[\beta(E_\lambda - E_\mu)]$ of the equilibrium probabilities for the system to be in $\lambda$ or $\mu$, for any $\lambda$ and $\mu$. For the RFIM, Dotsenko *et al.* have suggested a modification of this algorithm, in which the clusters are formed in exactly the same way, but instead of always flipping them over, the spins in the cluster are flipped with probability

$$A = \exp(-\beta \sum_i h_i s_i), \qquad (3)$$

where the sum is over only the spins in the cluster.* It is not hard to demonstrate that with this modification the algorithm satisfies the conditions of ergodicity and detailed balance for the random-field model. However, this is not a good algorithm for simulating the model in the difficult regime below the critical temperature for the following reason. As we decrease the temperature below $T_c$, larger ferromagnetic domains form, and the clusters produced by the Dotsenko algorithm become large. The argument of the exponential in Equation (3) is the sum over independent random variables with variance $\sigma^2$ and so will itself be a Gaussianly distributed random variable with variance $n\sigma^2$, where $n$ is the number of spins in the cluster. Thus, once the system becomes reasonably well equilibrated, and most domains are aligned with the prevailing direction of their local fields, acceptance ratios will fall to typical values on the order of $\exp(-\beta n \sigma^2)$ which is an exponentially small number for large clusters.

---

*It will be noticed that this acceptance ratio can become larger than one if enough of the spins are aligned opposite to their local fields. This problem can trivially be overcome by the standard procedure of dividing the acceptance ratios for a move to proceed in either direction by the larger of the two ratios.

The fundamental reason why this becomes a problem is that the clusters chosen by the Dotsenko algorithm are too large. The algorithm chooses its clusters according to the rules for the normal Ising model. But at a given temperature, the typical excitations in the normal Ising model are larger than those in the random-field case because random fields encourage the formation of local islands of spins that would not be energetically favored in the normal case. This is particularly evident when one considers the ferromagnetic "backbones" of spins which form in the two cases. These are the system-spanning percolating clusters of aligned spins that form as we enter the ferromagnetic phase. (In two dimensions, if there were a phase transition, there would be only one of these, breaking the up/down symmetry of the model as in the normal Ising case, but in three dimensions there can be more than one, pointing either up or down.) In the case of the Wolff algorithm for the normal Ising model, these backbones are flipped very often, since the chance of one of their spins being chosen as the seed for a cluster is large on account of their sheer extent. In the Dotsenko algorithm for the RFIM however, a backbone will *never* get flipped (in the limit of large lattice volume $N$) because the number of spins in the cluster is very large, and therefore the acceptance ratio is infinitesimal. (More accurately, it could be flipped *once* as the simulation equilibrates, should it happen to form pointing in the "wrong" direction, but after that it has an infinitesimal chance of flipping again.) Given the large amount of time it takes a computer to find one of these backbone clusters, such an infinitesimal acceptance ratio is a serious waste of CPU. Furthermore, although the RFIM only has system spanning clusters below its critical temperature $T_c^{(R)}$, the Dotsenko algorithm can generate a system-spanning cluster any time the temperature of the simulation falls below the critical temperature $T_c^{(I)}$ of the normal Ising model. Since $T_c^{(R)} < T_c^{(I)}$, this means that there is a range of temperatures above the transition for which the algorithm already has a very small acceptance ratio, even though the Metropolis algorithm has no difficulty simulating the model in this regime.

A better algorithm for simulating the RFIM would instead choose as a cluster only a part of an infinite backbone, and, if it should happen that the local fields in that particular part were at odds with the way the spins of the backbone were pointing, the acceptance ratio $A$ might well be moderately high, and the chances would be good that the cluster would be flipped. This suggests to us that we should try and find a modification of the Dotsenko algorithm in which the physical size of the clusters is limited to prevent the costly growth of very large ones whose chances of being flipped are negligible. A method for doing exactly this has been suggested by Barkema and Marko [25] in the context of simulations of spinodal decomposition using the conserved-order-parameter Ising model. In terms of the normal Ising model, their algorithm is as follows. Again the cluster starts with a single spin chosen at random, and again similarly-oriented spins neighboring those already in the cluster are added to it with probability $1 - \exp(-2\beta J)$, except that now they are only added if they are within some distance $r$ of the initial site from which the cluster was grown.[†] This indeed prevents the cluster from having a linear dimension larger than about $2r$, but at the same time it ruins the careful balance of probabilities in the Wolff algorithm which make it satisfy the condition of detailed balance. In particular, there are spins just outside the radius $r$ which in the Wolff algorithm would have a probability of $\exp(-2\beta J)$ of *not* being added to the cluster, but in the present algorithm have probability 1 of not being added. Thus the cluster we have formed is $\exp(2m\beta J)$ more likely to appear in our new algorithm than it would be in the thermally correct Wolff algorithm, where $m$ is the number of such spins lying just outside the radius $r$ which might have been added had they not been so unlucky about where they were situated. In order to rectify this "mistake" in the probability for formation of this cluster, we therefore introduce a new acceptance ratio

$$B = \exp(-2m\beta J) \qquad (4)$$

for the flipping of the cluster. This makes the flipping of any particular cluster exactly as likely as in the Wolff case, and therefore restores detailed balance.

To generalize this algorithm to the random-field case, we simply multiply the acceptance ratio $B$ by the previous one $A$, to get a new ratio

$$AB = \exp[-\beta(2mJ + \sum_i h_i s_i)]. \qquad (5)$$

This yields an algorithm—the Limited Cluster Flip or LCF algorithm—which obeys ergodicity and detailed balance, and has a hard limit of $r$ on the radii of clusters, which stops the very large time-wasting ones from forming. The LCF algorithm is a genuine improvement over both the Metropolis algorithm and the Dotsenko algorithm, having simultaneously smaller pinning problems than the former and a better acceptance ratio than the latter, but it still has difficulties. In fact, it suffers from exactly the same problem as the Metropolis algorithm, but at a slightly longer length scale. The algorithm tries to flip regions of size up to $2r$, but if such a region is lying on top of a number of local fields which are all pointing in the same direction, then the acceptance ratio (5) can be low, and the region is effectively "pinned". This can stop an island of spins from shrinking under the influence of surface tension just as effectively as the pinning of single spins in the earlier case. The algorithm *is* an improvement over the single-spin flip one because for islands whose size is less than $2r$ pinning is no longer a problem. However, as we cool below $T_c$ and the system coarsens with domains coalescing into larger domains, the typical domain size will always eventually become larger than $2r$, and then we are back where we started. This problem can also be understood using renormalization group arguments of the type employed, for instance, in Ref. [13], where spins are blocked to remove the small length-scale degrees of freedom and the model is mapped onto another RFIM with lattice parameter a factor $b$ larger and different values for $J$ and $\sigma$. If we repeat this process $n$ times such that $b^n = 2r$ we create a system which is still an RFIM but for which $2r$ is the lattice parameter, and for this system our algorithm is exactly equivalent to the Metropolis algorithm, and therefore has all the same problems (on length scales greater than $2r$).

So the problem is that, if we impose a particular length-scale $r$ on our cluster flipping algorithm, we have pinning problems whenever domain walls need to diffuse over distances greater than $2r$ in order to find other walls to annihilate with and thereby lower the free energy. But conversely, if we *don't* impose any particular length-scale on our clusters, as with the Dotsenko algorithm, we get infinitesimal acceptance ratios as the domain size increases and so the algorithm is hopelessly inefficient. Is there a way out of this dilemma? There seems to be in the following algorithm, which is the one we have employed for the simulations reported in this paper. We employ precisely the LCF algorithm as described above, but we *vary the length scale $r$* from each step to the next. In order to satisfy detailed balance this variation has to be independent of the state of the system, so we make the variation random by choosing a new value for the radius $r$ at random from a distribution $P(r)$ before each step. How should we choose this distribution? First, it should give more weight to small radii than to large ones; we know that if we pick a lot of large clusters, we will only waste time because their acceptance ratio will be low and they will not be flipped over. However, we need to pick *some* large clusters, to allow the domain walls to hop over pinned regions of any finite size. Second, we don't want to introduce any particular length-scale into the algorithm for the precisely the reasons described above,[‡] so distributions such as an exponential or Gaussian dis-

---

[†]The distance can be defined in various different ways. In their simulations, Barkema and Marko [25] defined it using a "Manhattan" formula in which the distance between two points $(x_1, y_1)$ and $(x_2, y_2)$ on a two-dimensional square lattice is $|x_1 - x_2| + |y_1 - y_2|$ (with the obvious generalization in higher dimensions). However, the exact definition of the distance is not important for the working of the algorithm.

[‡]In their simulations, Barkema and Marko used a length-scale that was tuned to the equilibrium correlation length at the temperature they were studying. However, this is not a shrewd idea in the present case, since the rapid decrease of the correlation length below $T_c$ means that the algorithm would become equivalent to the Metropolis algorithm again at temperatures only a little below the phase transition.

tribution are ruled out. The obvious scale-free candidate for the distribution is the power law

$$P(r) = Cr^{-\alpha}, \qquad (6)$$

with $\alpha > 0$. The constant $C$ is chosen to normalize the probability distribution to unity over the range of values $r$ can take in a particular simulation. The only remaining question is, what value should $\alpha$ take? Empirically we have found that the best results are obtained for values of $\alpha$ around 2 or a little larger. We can justify this if we assume that it is most efficient to spend a roughly equal amount of CPU time on clusters at all length scales. The time spent finding a cluster (and the time spent flipping it if the move is accepted) is approximately proportional to the number of spins in the cluster, as observed by Barkema and Marko [25], and therefore scales as $r^D$, where $D$ is the fractal dimension of the cluster. We would therefore expect that $\alpha = D$, and our observation that it actually takes a value around 2 suggests that the clusters in our algorithm have a fractal dimension of less than 3, and probably close to 2. As the temperature of the simulation decreases, we expect this dimension to increase, and that the optimal value of $\alpha$ will approach 3 as we go further and further below $T_c$.

Our algorithm has the advantage of a reasonably high acceptance ratio (typically between about five and ten per cent in the simulations reported here) as well as having a higher degree of immunity to domain wall pinning problems than any of the other algorithms. How well does it actually compare with other algorithms? In general it is not as efficient as the Metropolis algorithm in the regime well above the critical temperature, where the annealing away of domain walls is not an important equilibration process. But it really comes into its own in the region below $T_c$, which is the regime that has traditionally proved very hard to simulate. In Figure 1 we have compared our algorithm against the Metropolis algorithm for a typical system of $16 \times 16 \times 16$ spins on a three-dimensional cubic lattice. The parameters used where $J = \frac{1}{3}$ and $\sigma = 0.35$, which is well into the ferromagnetic region at moderately high disorder—the regime in which the new algorithm is expected to score most highly over the Metropolis algorithm. In Figure 1(a) we show the magnetization and internal energy (upper and lower pairs of curves respectively) as a function of time measured in Monte Carlo steps from the start of the simulation at $T = \infty$ until after both algorithms have equilibrated. The solid lines are the results from our cluster-flipping algorithm and the dashed ones are the results for the Metropolis algorithm. The equilibrium value of the magnetization is about 0.92 per spin and that of the internal energy is about $-0.89$, and it is clear from the figure that the new algorithm finds these equilibrium values considerably quicker than the Metropolis algorithm.

However, we should be wary of such comparisons, since the amount of CPU time taken to perform one Monte Carlo step in the new algorithm is considerably longer than that taken in the Metropolis algorithm, because of the complexity of the decisions involved.

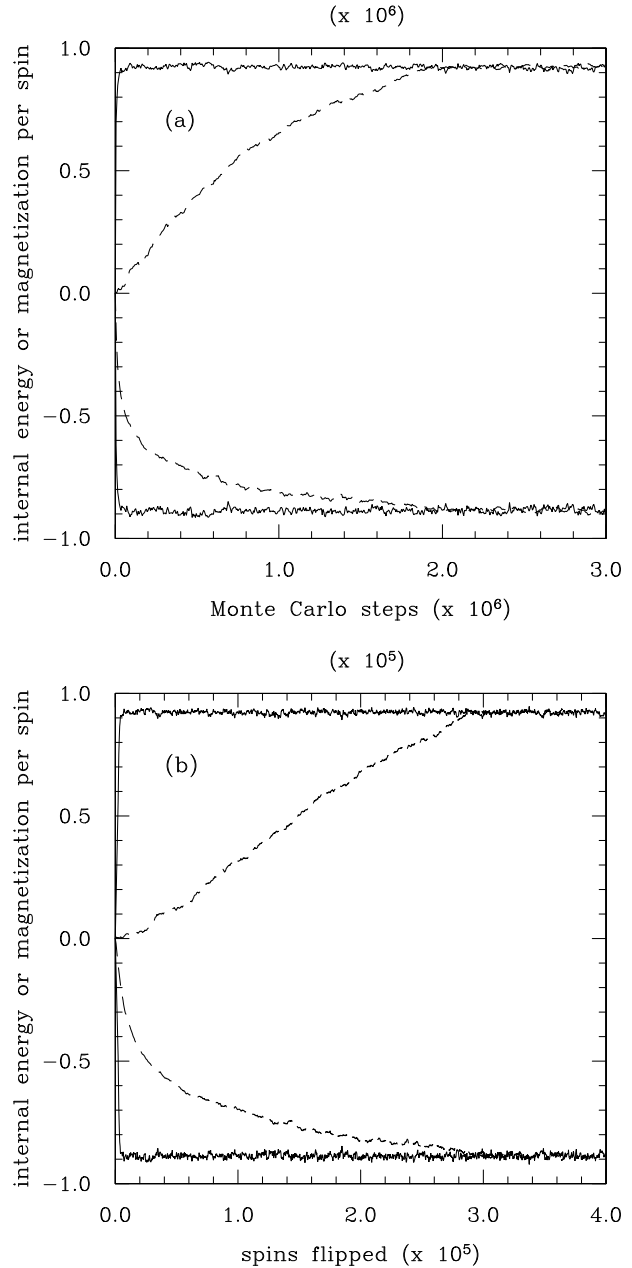

FIG. 1. Comparison of the rate of equilibration of the cluster-flipping algorithm proposed in the text (solid lines) and the Metropolis algorithm (dashed lines) for a three-dimensional RFIM system of $16 \times 16 \times 16$ spins on a cubic lattice, with $J = \frac{1}{3}$ and $\sigma = 0.35$. The upper pairs of lines in each figure are the magnetization of the system with time, and the lower ones the internal energy. In (a) the horizontal axis represents the number of Monte Carlo steps performed. In (b) it is the total number of spins flipped.

On the other hand of course, each step in the new algorithm can flip a number of spins, whereas the Metropolis algorithm flips at most one at each step. In fact, as is often the case with Monte Carlo simulations of simple models, if the code is reasonably efficient, the crucial question in comparing the speeds is how many random numbers you have to generate. In the Metropolis algorithm, four random numbers are generated per spin considered, whereas in the new algorithm an average of between one and three are generated per spin, depending on the temperature, giving the new algorithm the edge, particularly at low temperatures as it turns out. On the other hand the Metropolis algorithm has a somewhat higher acceptance ratio (by about a factor of two in the simulations presented here), so on balance it turns out that both algorithms generate about the same number of random numbers *per spin flipped*.

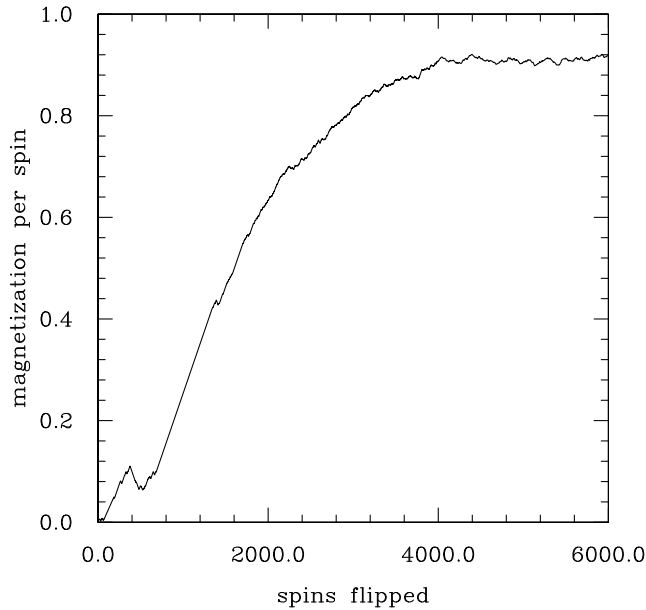

FIG. 2. A magnified plot of the magnetization for the cluster-flipping algorithm, and shows that for this system of $\sim 4000$ spins, equilibrium is reached after each spin has been flipped about once on average.

However, this comparison ignores the physical considerations which drove us to create the algorithm in the first place. The Metropolis algorithm may flip the same number of spins per random number generated on average as our cluster algorithm, but most of those flips just result in small random walks of pinned domain boundaries, which do little for the overall equilibration of the system. The cluster algorithm on the other hand is much more directed in the spins which it chooses to flip, and for this reason we hope it will be more efficient at finding the low-free-energy regions of the model's state space, and become more so as we go to lower and lower temperatures. Thus a truer comparison of the relative merits of the two algorithms would be to compare the rates of equilibration when time is measured in terms of the aggregate number of spins flipped by the algorithm. We do this in Figure 1(b). Clearly there is still a dramatic advantage to the cluster flipping algorithm (it is about a factor of 70 faster). In fact, as shown in Figure 2, the algorithm appears to find the equilibrium value of the magnetization after flipping roughly 4000 spins, which is about the best we can expect, given that the system being simulated only has 4000 spins in it. On average it flips each spin only once in coming to equilibrium. (As far as we are able to tell from the simulations we have performed, this rule of thumb generalizes to systems of other sizes, both larger and smaller, at least up to systems of 32 spins on a side, which is the largest size we have looked at. This suggests that the new algorithm's advantage over the exponentially slow Metropolis algorithm will improve quickly as the size of the system studied increases. The two orders of magnitude acceleration seen here for a $16^3$ system may be dwarfed by far more spectacular gains in larger systems.)

In practice the algorithm equilibrates very fast and has allowed us to perform substantially longer simulations of the RFIM than have previously been possible whilst using less computer time. The results presented in this paper are still for relatively small systems, because each system size has to be averaged over a large number of different realizations of the randomness in order to extract reliable results for the critical behavior. However, in the near future we hope to use a parallel implementation of our algorithm on the Cornell Theory Center's IBM SP–2 parallel computer to study significantly larger system sizes.

### III. REWEIGHTING SCHEME

Our plan then is to use our new Monte Carlo algorithm to simulate the RFIM in three dimensions in the region of the phase transition between the para- and ferromagnetic states for a variety of different sizes of system, and then use finite-size scaling to extrapolate the critical behavior of the infinite system from these results. The pinning effect of the local fields, which is the main thing slowing our simulations down, becomes more pronounced as we increase the width $\sigma$ of the distribution from which the fields are drawn, so in order to save time and improve the accuracy of our simulation, we would like to keep this width as small as possible. On the other hand, the size of the scaling regime close to the critical temperature in which the model has behavior representative of the critical properties of the RFIM gets smaller as $\sigma$ is decreased, as discussed below in Section IV. What we would like to do therefore, is to perform simulations for a range of different temperatures $T$ (or coupling constants $J$) and randomnesses $\sigma$, and then apply scaling methods to estimate the size of the scaling regime and extract the best results we can from these data. To this end we have employed a generalization of the "histogram method" of Ferrenberg and Swendsen [26], which is a reweighting scheme which returns the best estimate of the partition function of a system given a particular set of measurements from one or more Monte Carlo simulations. Our generalization allows us to perform simulations at a small number of different temperatures and randomnesses, and then make an estimate of the partition function at any intervening temperature or randomness. In addition, the method gives results that are of higher statistical quality than the raw data from the simulations: the errors are smaller, and the fluctuations from one data point to the next are decreased.

The method is a straightforward generalization of the one proposed in Ref. [26]. It can be applied to any simulation of a Hamiltonian system in which the Hamiltonian can be written as the sum of a number of terms each consisting of an independent coupling constant $J^{(n)}$ times an interaction energy $E^{(n)}$:

$$H = \sum_n J^{(n)} E^{(n)}. \qquad (7)$$

The Hamiltonian for the RFIM can be written in this form with two terms in which

$$J^{(1)} = J, \qquad E^{(1)} = \sum_{\langle ij \rangle} s_i s_j,$$
$$J^{(2)} = \sigma, \qquad E^{(2)} = \sum_i k_i s_i, \qquad (8)$$

where

$$k_i \equiv \frac{h_i}{\sigma}, \qquad (9)$$

which are the magnitudes of the random fields, normalized so that their variance is unity. Writing the Hamiltonian in this way means that if we know the values of the interaction energies $E^{(1)}$ and $E^{(2)}$ for any particular state of the system we can calculate $H$ for that state for any $J$ and $\sigma$ simply by plugging the values into Equation (7).

Now suppose we have performed a number of different Monte Carlo simulations for a range of different values $J_i^{(n)}$ of the parameters $J^{(n)}$. For each simulation (denoted by the subscript $i$) we allow the system time to equilibrate, and then record a set of values of the interaction energies, which we denote $E_{ij}^{(n)}$ with $j$ running from 1 to however many values we record. The values do not have to be taken at successive Monte Carlo steps. In fact it is most efficient to take them at intervals of one correlation time. However, if they are taken at intervals more or less frequent than that, the method will still work.

We write our estimate of the partition function $Z_q$ of the system at the values $J_q^{(n)}$ used in the $q^\text{th}$ run, in the form:

$$Z_q = \sum_{ij} a_{ij} \exp(-\beta \sum_n J_q^{(n)} E_{ij}^{(m)}). \qquad (10)$$

The $a_{ij}$ are new quantities—weights—which we introduce and whose values we are going to choose in order to make $Z_q$ as good an estimate of the true partition function as we can. There is one such weight for each microstate $(i,j)$ sampled in one of the runs. The sum in the exponential is just the value the Hamiltonian (7) would take if the system were in the state $(i,j)$ and the coupling constants took the values $J_q^{(n)}$. Thus, if all the $a_{ij}$ are set equal to unity, Equation (10) is just a straightforward estimate of the partition function, using all the states of the system that we know about and none of the ones that we don't.

However, this is not the best possible estimate of the partition function. Far from it, in fact, since it can include many states $(i,j)$ which are very unlikely to be sampled in a Monte Carlo run at the given values of the $J^{(n)}$, but fails to include many others which are relatively likely but didn't get sampled simply because the runs we do only have time to sample a very small fraction of the possible states. We can improve our estimate considerably by adjusting the weights $a_{ij}$ to reflect these considerations. If a particular state $(i,j)$ is relatively likely to be sampled in a particular run, then other states with energies nearby (if there are any) should be likely too, and even if those states don't actually get sampled in the particular simulation we perform, their contribution to the partition function can be approximated by increasing the value of the corresponding weight $a_{ij}$. In a sense, $a_{ij}$ reflects our estimate of the density of states near state $(i,j)$. With a suitable choice of all the weights, we should be able to get a good estimate of the partition function given all the Monte Carlo data we have from the different runs.

How exactly then do we choose the weights? Well, we want to choose them in a way which reflects the fact that the states $(i,j)$ through which the system passes during run $i$ are on average much more likely to occur during that run than most states. We define the "quality function" $Q$, which is the probability that these states (or ones close to them in energy) will be the ones that actually do crop up in the simulation:

$$Q = \prod_{kl} \frac{a_{kl} \exp(-\beta \sum_n J_k^{(n)} E_{kl}^{(n)})}{Z_k}. \qquad (11)$$

If we maximize this by adjusting the weights, we can find the values of the $a_{ij}$ for which this set of states is the *most likely* to occur, and this is what we use as our criterion for choosing the weights. Differentiating to find the maximum, we get

$$\frac{1}{a_{ij}} = \sum_k \frac{N_k}{Z_k} \exp(-\beta \sum_n J_k^{(n)} E_{ij}^{(n)}), \qquad (12)$$

where $N_k$ is the number of samples taken during the $k^\text{th}$ run. Substituting this back into Equation (10) we get an expression for $Z_q$ in terms of the complete set of $\{Z_k\}$. Starting with a sensible guess for each of the $Z_k$ (setting $a_{ij} = 1$ for all $i,j$ is a good choice) and iterating, we converge on our best value for $Z$, usually quite rapidly.[§] Then we can substitute these values for the $Z_k$ back into Equation (12) one last time, to get the best values for $a_{ij}$. These we can then use to evaluate the best estimate of any observable $X$, according to

$$\langle X \rangle = \frac{1}{Z} \sum_{ij} X_{ij} a_{ij} \exp(-\beta \sum_n J^{(n)} E_{ij}^{(n)}), \quad (13)$$

for any set of values $J^{(n)}$ we like. (They do not have to be any of the values at which the simulations were performed. They can be in between those values giving a best guess for the interpolation between data points, or they can be outside the range of those values, giving an extrapolation.) Here $X_{ij}$ is the value of $X$ in the state denoted by $(i,j)$. Thus for example $X_{ij}$ could be the sum $\sum_m s_m$ of all the spins in our RFIM system in state $(i,j)$, in which case $\langle X \rangle$ would be the magnetization $m$. Or we could set $X_{ij}$ equal to the Hamiltonian in that state, in which case $\langle X \rangle$ becomes the internal energy.

Figure 3 illustrates the working of the method. The discrete points are raw data for the magnetic susceptibility from Monte Carlo simulations on a small RFIM system, taken at three different randomnesses $\sigma$ and a variety of different temperatures. The family of solid lines are the results from the reweighting method over the same range of randomnesses, and illustrate how the method allows us to interpolate between the data points in both randomness and spin-spin coupling. What are not shown in the figure are the errors on the reweighted curves, which are considerably smaller than the errors on the raw data, an improvement which turns out to be very helpful when we come to extract critical exponents from our simulation results.

## IV. FINITE-SIZE SCALING ANALYSIS

We have performed simulations of RFIM systems at a variety of values of the coupling constant $J$ and the randomness $\sigma$ near the phase boundary between ferromagnetic and paramagnetic phases. In all the simulations we set $\beta = 1$. Starting from the high-temperature state of random uncorrelated spins we allow an initial time for the system to cool to equilibrium and then we record the value of the magnetization per spin $m$ at regular intervals, along with the values of the two interaction energies $E^{(1)}$ and $E^{(2)}$ defined in Equation (8). Armed with these measurements we use the reweighting scheme described in the last section to calculate curves for $\langle m(J) \rangle$ and $\langle m^2(J) \rangle$ for a variety of values of $\sigma$. We then use these to calculate the magnetic susceptibility per spin $\chi$ and the disconnected susceptibility $\chi_{\text{dis}}$ from

$$\chi = \langle m^2 \rangle - \langle m \rangle^2, \quad (14)$$
$$\chi_{\text{dis}} = \langle m^2 \rangle. \quad (15)$$

However, these results show large fluctuations from one realization of the random fields to another, so we repeat the entire calculation a large number of times (150 in the present case) and average over all of them to get mean values $[m]_{\text{av}}$, $[\chi]_{\text{av}}$, and $[\chi_{\text{dis}}]_{\text{av}}$ for $m$, $\chi$, and $\chi_{\text{dis}}$. The errors in our results are primarily an indication of the size of the sample-to-sample fluctuations in the measured quantities, since these fluctuations turn out to be larger than the statistical errors introduced by the Monte Carlo method.

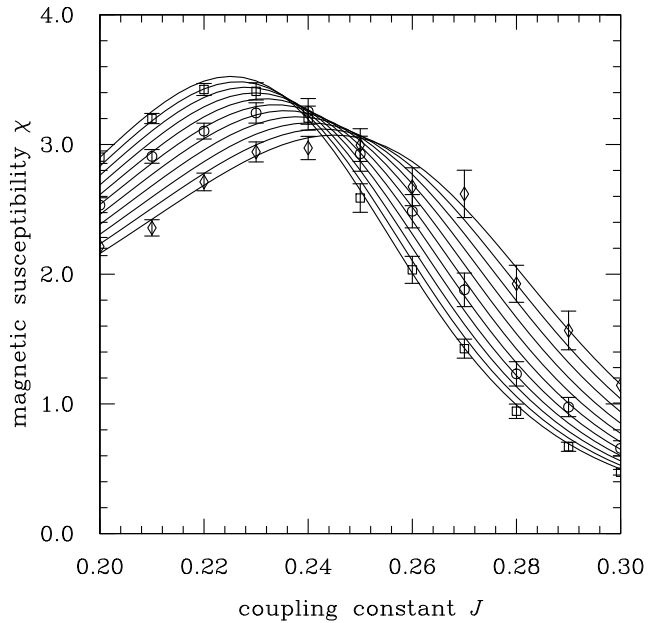

FIG. 3. Illustration of the working of the reweighting scheme described in Section III. The data points with error bars represent the raw data for the magnetic susceptibility for three different values of the randomness $\sigma$. The family of solid lines are the best estimate of the susceptibility as a function of coupling constant $J$ for a variety of values of $\sigma$ covering the same range, calculated from the reweighted partition function. Note that the reweighting allows us to interpolate in both coupling and randomness.

We use our results for $[m]_{\text{av}}$, $[\chi]_{\text{av}}$, and $[\chi_{\text{dis}}]_{\text{av}}$ as the basis for a finite-size scaling extrapolation of the values for the critical exponents of the RFIM. We repeat our calculations for systems of a number of different sizes $L^3$ and, following Rieger and Young [17], we assume the scaling forms

$$[m]_{\mathrm{av}} = L^{-\beta/\nu} \widetilde{m}(L^{1/\nu} t), \qquad (16)$$

$$[\chi]_{\mathrm{av}} = L^{\gamma/\nu} \widetilde{\chi}(L^{1/\nu} t), \qquad (17)$$

$$[\chi_{\mathrm{dis}}]_{\mathrm{av}} = L^{\bar{\gamma}/\nu} \widetilde{\chi}_{\mathrm{dis}}(L^{1/\nu} t), \qquad (18)$$

for each set of curves for a given value of $\sigma$. The functions $\widetilde{m}(x)$, $\widetilde{\chi}(x)$, and $\widetilde{\chi}_{\mathrm{dis}}(x)$, are universal functions of the scaling variable $x \equiv L^{1/\nu} t$ which should be independent of the values of the microscopic parameters $J$ and $\sigma$. The variable $t$ is defined as

$$t \equiv \frac{J_c - J}{J_c}, \qquad (19)$$

where $J_c$ is the critical value of $J$ for a given randomness, which is not expected to be a universal quantity. The exponents $\beta$, $\gamma$, $\bar{\gamma}$, and $\nu$ are the usual critical exponents for the model, defined by the behavior of the RFIM in the thermodynamic limit near criticality:

$$m \sim t^\beta, \quad \chi \sim t^{-\gamma}, \quad \chi_{\mathrm{dis}} \sim t^{-\bar{\gamma}}, \quad \xi \sim t^{-\nu}. \qquad (20)$$

---

[§]In fact the speed with which the iteration converges seems to depend on how well the samples that we have taken cover the important ranges of energy. If we try and get the method to converge for values of $J^{(n)}$ for which the energies of the most likely states fall far away from any of the energies sampled during our Monte Carlo runs, convergence will be slow, and even after convergence the method will not give a very good estimate of the partition function. For this reason, the method is not suitable for extrapolating far beyond the regime in which the simulations were performed, or for interpolating between simulations with very different values of the parameters $J^{(n)}$.

($\xi$ is the correlation length.)

For any particular value of $\sigma$, we can use the scaling forms (16), (17), and (18) to extract values for the critical exponents by rearranging them to give the scaling functions $\widetilde{m}(x)$, $\widetilde{\chi}(x)$, and $\widetilde{\chi}_{\mathrm{dis}}(x)$ and substituting in the values from our simulations for various sizes of system. Since the scaling functions are universal, the resulting curves should fall, or "collapse", on top one another if we use the correct values of the exponents $\beta$, $\gamma$, $\bar{\gamma}$, and $\nu$, and the critical coupling $J_c$ in the calculation. So, we simply adjust the values of these quantities to optimize the extent to which the curves for different sizes of system match. The cleanness of the data from the reweighting method has allowed us to use a computational method to perform this optimization, in which the variance across the several different system sizes of the calculated values of the scaling functions was integrated over a range of the scaling variable $x$ from $x_c - \Delta x$ to $x_c + \Delta x$, where $x_c$ and $\Delta x$ are some quantities which we choose according to criteria discussed below. Minimizing this integral using a standard "simplex" minimization algorithm yields the best values for the critical exponents and the critical coupling for given $\sigma$ and $\Delta x$. Figure 4 shows an example of a collapse of our data for the magnetization $m$ for four different sizes of system $L = 4, 6, 8, 12$ using the values of the exponents $\beta$ and $\nu$ given in the next section.

Since the critical exponents are expected to be universal quantities, they should take the same values regardless of what value of $\sigma$ we perform the analysis for. In fact, we can make best use of all the data we have for different values of the randomness by *simultaneously* minimizing the variance of the scaling functions for all the different randomnesses. Each value of $\sigma$ should have its own critical coupling $J_c$, since this is a non-universal quantity, and so as a bonus, this method gives us a picture of the phase boundary $J_c(\sigma)$ for the model, as well as more accurate critical exponents.

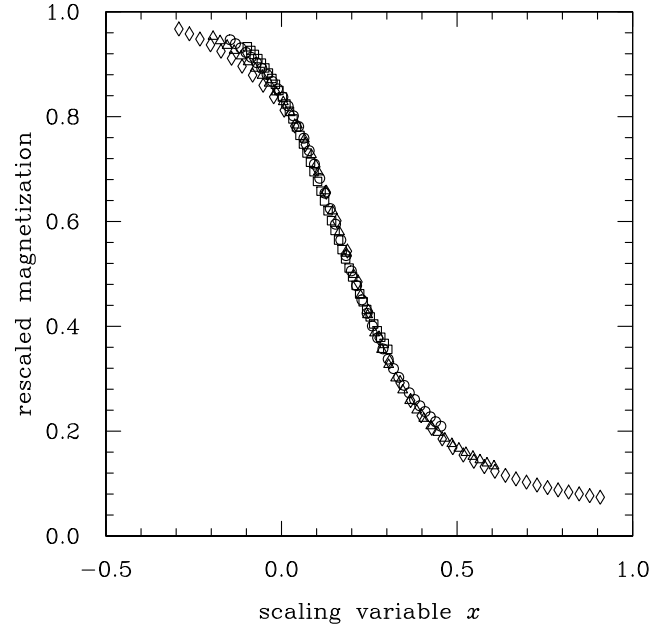

FIG. 4. Collapse of magnetization data from our simulations achieved using the scaling form given in Equation (16), with values of $\beta = 0.06$ and $\nu = 1.02$ for the exponents. The randomness was $\sigma = 0.35$ for these particular curves, and the critical coupling was taken to be $J_c = 0.27$.

The question still remains, how do we choose the value of the position $x_c$ and range $\Delta x$ of the scaling variable, over which the collapse is calculated? In the case of the susceptibility, $x_c$ is chosen to be the value at which the scaling function $\widetilde{\chi}(x)$ peaks, since this is the point of largest fluctuations in the magnetization, which we take as a indication of the position of the phase transition. In the case of $m$ and $\chi_{\mathrm{dis}}$, $x_c$ is taken to be the point of steepest gradient in the scaling function.

The choice of $\Delta x$ turns out to be crucial in extracting good results from our method. The regime in which the simulations display critical behavior typical of the RFIM is limited to values of the parameters $J$ and $\sigma$ close to the phase boundary. As demonstrated by scaling arguments such as those of Bray and Moore [12], the system displays behavior typical of the critical regime in a normal Ising model once we get sufficiently far away from the transition. Thus if $\Delta x$ is made too large, our method will simply measure the normal Ising model exponents. On the other hand, if $\Delta x$ is made very small, we end up throwing away most of the data for our scaling functions, and hence increasing the error bars on our values for the critical exponents. As a compromise therefore, we have adopted the method illustrated in Figure 5. We perform the minimization and collapse for a range of different values of $\Delta x$, from the very large ones which return values typical of the normal Ising model, to the very small ones, which are well inside the RFIM scaling regime, but which give large error bars on the quantities of interest. In the figure we have plotted the exponents $1/\nu$ and $\beta/\nu$ which appear in Equation (16) as functions of $\Delta x$, with the calculated error bars, and then performed a weighted extrapolation to the limit $\Delta x \to 0$ to calculate the final values for these exponents, by fitting the data to a simple quadratic. An alternative method employed by Rieger and Young [17,18] uses only the value of the scaling function at the value of $x$ which maximizes $\widetilde{C}$, the scaling function for the specific heat (which we have not calculated here). This method is roughly equivalent to our performing our collapse only for $\Delta x = 0$.** Although this is guaranteed to avoid crossing over into the normal Ising model regime, it also effectively throws away a lot of the simulation data, and therefore gives larger errors on the values of the critical exponents.

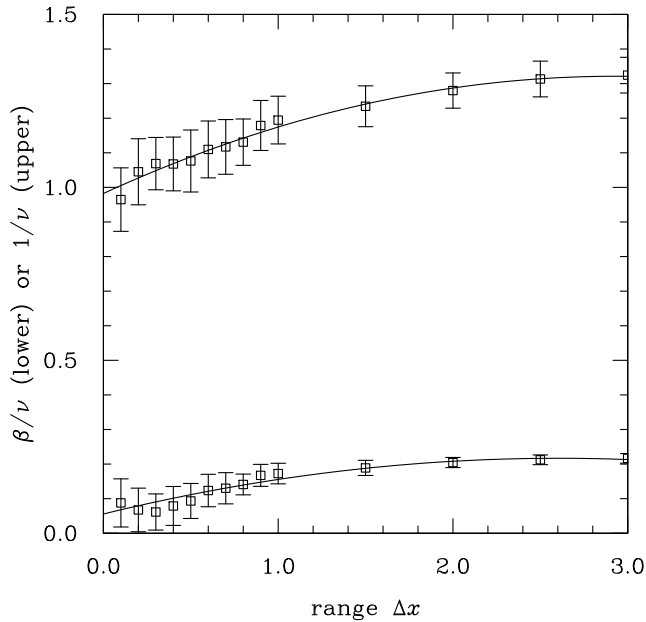

FIG. 5. Extrapolation of the exponents $1/\nu$ (upper curve) and $\beta/\nu$ (lower curve) to the limit $\Delta x \to 0$ as described in Section IV.

---

**To be strictly equivalent, we would have to perform our collapse in a range centered around the maximum of the scaling function for the specific heat. But since the difference between the position of this maximum and the positions $x_c$ that we have used in our collapses are much smaller than the smallest values of $\Delta x$, we can ignore this point.

## V. RESULTS

For each system simulated, we fixed the inverse temperature $\beta$ at 1, and performed Monte Carlo runs at eleven different values of the coupling constant $J$ between 0.2 and 0.3, for each of three different values $\sigma = 0.25$, 0.35, and 0.45 of the randomness, giving 33 runs on each system in all. We studied 150 different realizations of the randomness in systems of linear dimension $L = 4, 6, 8, 12$, and 16, for a total of twenty-five thousand simulations. (We have also performed a smaller number of runs on systems of size $L = 20, 24$ and 32. However, these results, along with those for $L = 16$ have not been used in most of the calculations presented here, since the errors in the measured quantites due to sample-to-sample fluctuations get worse as we go to larger systems because of the non-self-averaging nature of the RFIM [27].) Each simulation consisted of starting the system in a random-spin $T = \infty$ configuration, cooling the system to equilibrium using our algorithm, and simulating for about a further 10 equilibration times, during which approximately 200 samples of the system variables were recorded, as described above. The reweighting analysis was performed afterwards as a separate stage in the calculation. In order to perform such a large number of long Monte Carlo simulations, we made use of the parallel computing power provided by the IBM SP–1 and SP–2 computers at the Cornell Theory Center. Most of the calculations presented here were performed using a parallel version of our code on 75 processors of the SP–2 computer. The reweighted curves for $m$, $\chi$, and $\chi_{\rm dis}$ were averaged over the many different samples, the scaling collapse performed by minimizing the variance of the scaling functions with respect to the critical exponents as described above, and the extrapolation to the limit $\Delta x \to 0$ performed using a weighted least-squares fit. The errors on the exponents and the critical coupling $J_c$ were calculated using a "bootstrap" resampling method [28].

The RFIM is believed to have possibly two, but probably three, independent critical exponents (see Ref. [1]). Our calculations of $m$, $\chi$, and $\chi_{\rm dis}$ pinpoint the values of four of the exponents, and allow us to evaluate any others using scaling relations such as those proposed by Bray and Moore [12]. From the calculations described above, our best estimates of the exponent ratios $\beta/\nu$, $\gamma/\nu$, and $\bar{\gamma}/\nu$ appearing in the scaling relations (16), (17), and (18) are:

$$\frac{\beta}{\nu} = 0.056 \pm 0.065,$$
$$\frac{\gamma}{\nu} = 1.851 \pm 0.067,$$
$$\frac{\bar{\gamma}}{\nu} = 2.843 \pm 0.066. \qquad (21)$$

The three data collapses also give us three measurements of the value of $\nu$. Combining these, our best estimate for $\nu$ is

$$\nu = 1.022 \pm 0.057. \tag{22}$$

These figures are in agreement with previous estimates of the same quantities by a variety of methods, with errors better than or comparable to those studies. In particular we agree with the figures given in previous Monte Carlo studies and while our errors are as good as or better than the most accurate of these studies [18], our computational effort has been considerably smaller. A detailed comparison with previous work is given in Table 1.

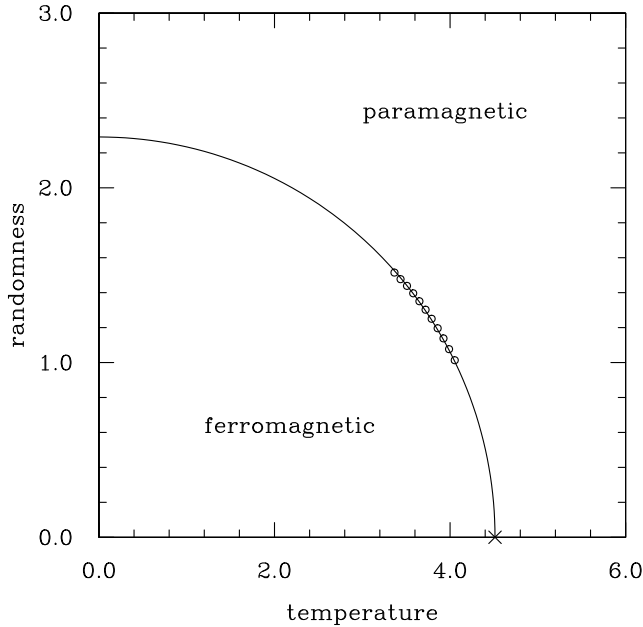

FIG. 6. Values of $T_c$ from the finite-size scaling analysis of Section IV (circles), the known $T_c$ for the normal Ising model (cross), and a simple extrapolation of the phase boundary to $T = 0$ by fitting to an ellipse. The estimated critical value $\sigma_c$ of the randomness above which no phase transition takes place is $2.3 \pm 0.2$.

In addition, Dahmen and Sethna [6] have suggested on the basis of arguments similar to those of Parisi and Sourlas [2] that within perturbation theory the exponents for the RFIM should be the same as those describing the hysteretic phase transition of the out-of-equilibrium, zero-temperature RFIM with varying external field [29]. (There may be non-perturbative corrections which ultimately mean that the exponents will be different, though Maritan et al. [30] have argued that the exponents may actually take the same values for both models.) Simulations performed by Perković [31] indicate that for that model the exponents equivalent to our $\beta$ and $\gamma$ have the values $0.04 \pm 0.04$ and $1.8 \pm 0.4$ respectively in three dimensions, which are in agreement with our figures. The exponent equivalent to our $\nu$ takes the value $1.4 \pm 0.2$, which is harder to reconcile with our figure of $\nu = 1.02 \pm 0.06$, but certainly does not rule out the possibility that the two are in fact equal.

The exponents given above should also satisfy the scaling relation [1]

$$2\frac{\beta}{\nu} + \frac{\bar{\gamma}}{\nu} = d. \tag{23}$$

where $d$ is the number of spatial dimensions, which is 3 in this case. Our value for this combination of exponents is $2.96 \pm 0.20$, clearly in agreement with the scaling relation.

Our finite-size scaling collapse also provides us with an estimate of the critical value $J_c$ of the coupling constant for each value of the randomness. We have used these values to calculate the critical *temperature* of an RFIM with randomness $\sigma$ and coupling constant $J = 1$, and plotted the resulting phase diagram in Figure 6. The circles represent our results, and the cross is the known transition temperature of the normal ($\sigma = 0$) Ising model in three dimensions. The solid line is a simple ellipse, with parameters chosen to best pass through the data points. This gives us a crude extrapolation of the phase boundary to the limit $T = 0$, yielding an approximate figure of

$$\sigma_c = 2.3 \pm 0.2 \tag{24}$$

for the critical randomness above which there is no phase transition at any temperature. This compares favorably with the estimate of $\sigma_c = 2.35$ calculated by Ogielski [32] using an algorithm which finds the ground state of RFIM systems.

## VI. CONCLUSIONS

We have designed and implemented a cluster-flipping Monte Carlo algorithm capable of equilibrating RFIM systems of moderate size significantly faster than either the Metropolis algorithm (which has been used exclusively in previous Monte Carlo studies of the model) or the algorithm suggested by Dotsenko et al. [23]. Using a parallel implementation of this algorithm in combination with a generalization of the histogram reweighting scheme of Ferrenberg and Swendsen [26], we have conducted extensive simulations to extract accurate data for the magnetization of RFIM systems of a variety of sizes in three dimensions with values of the coupling and randomness near the phase boundary between paramagnetic and ferromagnetic states. Using these data, we have performed a finite-size scaling analysis of the model and extracted results for the critical exponents of the model and for its critical temperature as a function of randomness. These results are in agreement with and of similar or better accuracy than the best available figures using competing methods.

The particular combination of techniques used in this study—the cluster-flipping algorithm, the reduction of errors and interpolation using a reweighting scheme, the finite-size scaling analysis and extrapolation to calculate

exponents—appears to be a particularly successful strategy for overcoming the unusual technical difficulties involved in the Monte Carlo simulation of the random-field Ising model. Preparations are at present under way to use these techniques in a larger study to provide a definitive calculation of the critical properties of this important model.

## VII. ACKNOWLEDGEMENTS

The authors would like to thank J. P. Sethna for a number of interesting and useful discussions concerning the calculations described here. This work was funded in part by the NSF under grant numbers ASC-9404936, ASC-9310244, and DMR-9121654, and by the Cornell Theory Center. The calculations were performed using computing facilities at the Cornell Theory Center.

| exp | value | Rieger[a] | Ogielski[b] | GAAHS[c] | DSY[d] |
|---|---|---|---|---|---|
| $\nu$ | $1.02 \pm 0.06$ | $1.1 \pm 0.2$ | $1.0 \pm 0.1$ | — | $\simeq 1.4$ |
| $\beta$ | $0.06 \pm 0.07$ | $0.00 \pm 0.05$ | $\simeq 0.05$ | — | $\simeq 0$ |
| $\gamma$ | $1.89 \pm 0.17$ | $1.7 \pm 0.2$ | — | $2.1 \pm 0.2$ | $1.9 - 2.2$ |
| $\bar{\gamma}$ | $2.91 \pm 0.23$ | $3.3 \pm 0.6$ | $\simeq 2.9$ | — | $\simeq 3.0$ |

TABLE I. Comparison of our results for the critical exponents with figures from previous calculations. [a]H. Rieger, Metropolis Monte Carlo calculation, Ref. [18]. [b]A. Ogielski, numerical ground-state calculations, Ref. [32]. [c]M. Gofman, J. Adler, A. Aharony, A. B. Harris, and M. Schwartz, high-temperature series, Ref. [33]. [d]I. Dayan, M. Schwartz, and A. P. Young, real-space renormalization group, Ref. [14].